\documentclass[dvips]{article}
\usepackage{icrctc07}

\title{Cosmic Rays and Global Warming}
\shorttitle{CR and Global Warming}

\authors{T. Sloan $^{1}$, A.W.Wolfendale$^{2}$ }
\shortauthors{T. Sloan and et al.}
\afiliations{$^1$Physics Department, University of Lancaster, Lancaster, UK\\
$^2$Physics Department, Durham University, Durham, UK\\}
\email{t.sloan@lancaster.ac.uk}

\abstract{It has been claimed by others that observed temporal
correlations of terrestrial cloud cover with `the cosmic ray
intensity' are causal.  The possibility arises, therefore, of a
connection between cosmic rays and Global Warming.  If true, the
implications would be very great.\\
We have examined this claim to look for evidence to corroborate 
it. So far we have not found any and so our tentative conclusions 
are to doubt it. Such correlations as appear
are more likely to be due to the small variations in solar irradiance,
which, of course, correlate with cosmic rays. We
estimate that less than 15$\%$ of the 11-year cycle warming variations are
due to cosmic rays and less than 2$\%$ of the warming over the last 35 years
is due to this cause.}

\begin{document}
\maketitle
\section{Introduction}

A phenomenon with strong politico-social implications is the
apparent correlation of cosmic ray (CR) intensity with low level
cloud cover (CC) - and thereby with mean global temperature 
\cite{MS1,MS3}.  Insofar as there is a possible
link of clouds with CR via ionization the correlation cannot be
dismissed out of hand.  It is not sufficient to say that the
energy content of CR is minute in comparison with solar irradiance
(SI) and therefore the effect must not be causal; the atmosphere
is a highly complex system with subtle properties and the idea
must be tested.
\section{CR and Cloud Cover}
In reference \cite{MS1} a correlation was demonstrated 
for `low clouds' ($<$3.2 km in altitude) between the 
changes in CC (the CC anomaly), and CR count rate as 
measured by the Hunacayo 
neutron monitor (see figure 1 of reference \cite{MS1}).   
The CC anomaly was derived from the ISCCP D2 analysis using 
the infrared data\cite{ISCCP}. It was then implied in \cite{MS1} 
that  the CR variation caused that in the CC. Since this may not be the 
case if both effects are correlated to a third variable, it is 
prudent to look for further evidence of such a causal connection. \\
The first problem that arises is that the correlation is 
absent in the data for two other atmospheric depths: `middle
levels' (3.2 - 6.5 km) and `high levels' ($>$6.5 km altitude). 
This result is surprising in view of the fact that the CR
ionization (mainly from muons and electrons) increases with
height.  Specifically the rate of production of ions, in
cm$^{-3}$s$^{-1}$, for the 3 levels is estimated to be: 
high, 130(50); middle,
30(13) and low, 4(3) where the values are for $60^{\circ}$N (the
equator).\\
A possibility, and one needed by the proponents of the CR-CC
causal connection, is that the efficiency of the conversion from CR ions to
cloud droplets (presumably by way of aerosols) falls with
increasing height above sea level.  Such behaviour cannot be ruled
out but seems rather ad hoc.  The implication would be that even
in the low region the efficiency in not 100\% and, as will be
shown, there are already problems in this respect.\\
\section{Efficiency of ions for cloud production}
An important aspect is that of the likely efficiency of CR ions
for initiating cloud droplets and we start by estimating the
density of cloud droplets that could be produced by CR (muons) at
the lowest level.  With a rate of ions, of 4 cm$^{-3}$ s$^{-1}$ and
assuming they all give `small ions' the rate of production of
small ions will be the same, giving, for a mean lifetime of 50 s
\cite{Chalmers} 200 cm$^{-3}$. To produce significant nucleation
rates much larger ion densities than this were required \cite{MS2}. 
Hence the ionisation rate in CR could be too
small to produce significant numbers of water droplets such as
would be necessary in a cloud.\\
\section{Latitude dependence of `the effect'}
\begin{figure}
\vspace*{-3mm}
\begin{center}
\includegraphics [width=0.5\textwidth]{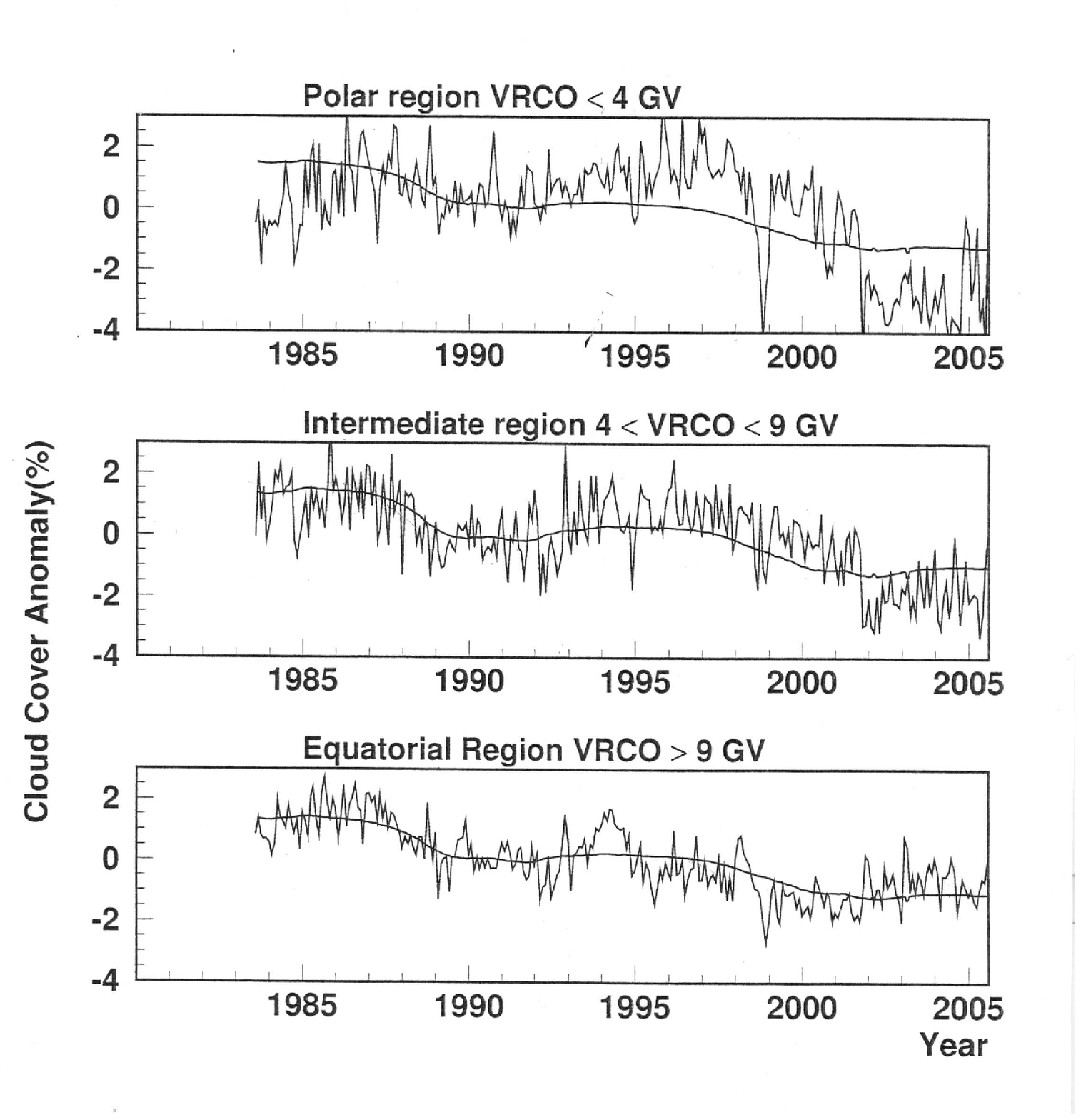}
\end{center}
\vspace*{-8mm}
\caption{CC anomaly as a function of time for various 
ranges of VRCO. The smooth curve shows a fit of the 
monthly mean of the daily sun spot (SS) rate with an 
assumed linearly falling background. VRCO is the vertical 
cut-off rigidity. The SSN is anticorrelated with 
the CR count rate with a lead time of several months.}\label{fig2}
\end{figure}

It is well known that the magnitude of the CR time variation,  
due to the 11 year solar cycle, varies with latitude.  More
accurately, it is a function of the vertical rigidity cut-off 
(VRCO), the reason being the effect of the geomagnetic field
deflecting away more low energy particles as the geomagnetic
equator (highest VRCO) is approached.  Since this variation falls
with increasing primary CR energy, the solar modulation is most severe
in the polar regions. Hence one would expect larger changes in CC 
in the polar regions than at the Equator.  Furthermore it is 
known that there is a delay of 6-14 months between the 
decrease in the CR intensity and the increase in the sun spot (SS) 
number with even numbered solar cycles showing smaller delays than 
the odd numbered \cite{Kudela}. Note that the CR count rate is 
anticorrelated to the SS number. \\
We have studied this effect in some detail by plotting the CC for
different VRCO bands and the results are given in Figure \ref{fig2}. 
The smooth curves in figure \ref{fig2} show the best fit of the 
CC anomaly to the mean daily SS number (inverted) with a linearly 
changing background. We observe the same dip in CC as seen in \cite{MS1} 
between the years 1985 and 1995. However, the 
expected rise in amplitude of this dip with increasing VRCO 
is not apparent. Furthermore, the dip 
in CC seen in solar cycle 22 (peaking in 1990) is not  
evident in solar cycle 23 (peaking in 2000) except in the equatorial 
region (high VRCO) where the solar modulation is least. To investigate 
this effect further and to check that the above result was not due 
to a latitude dependent 'efficiency'  the CC was determined in three 
strips of latitude and the amplitude of the dip in solar cycle 22 
was measured for each as a function of VRCO.  
This was achieved by fitting the SS number variation to the observed 
CC again with a linearly decreasing background. The delay between 
the onset of the dip in CC and that of the 
SS number was also a free parameter in the fit. 
Figure \ref{fig3} (upper panel) shows that the amplitude of the dip 
appears to be constant with VRCO 
rather than changing in an analogous manner to the 
observed CR modulation \cite{Braun}. The measured value of 
the delay between the onset of the dip and the change in SS number  
fluctuates randomly rather than concentrates around a fixed delay 
(expected to be $-7$ months for solar cycle 22). Hence there is 
an imperfect time correlation between the start of the
dip and the change in the CR rate. Thus the data in figure \ref{fig3} 
do not corroborate the claim of a causal connection between CR 
and CC.\\
To identify the parts of the distribution in 
the upper panel of figure \ref{fig3} which correlate with 
the CR modulation a fit was performed of the shape 
of the neutron modulation curve (the correlated part) and  
a constant term (the uncorrelated part) to the measurements. This fit 
showed that less than $15\%$ of the distribution at 95$\%$ confidence 
level belonged to the correlated part.\\   
\section{The Chernobyl Nuclear Accident} 
On the 26 April 1986 there was a nuclear reactor accident  
at Chernobyl (51.4$^\circ$ N 30.1$^\circ$ E) which released 
large amounts of radioactivity into the atmosphere.  
A correlation between CC and ionization from the radioactivity  
would be expected to produce an increase in CC in 
the vicinity of Chernobyl following the accident 
if \cite{MS1} were correct. 
Figure \ref{fig4} shows the CC anomaly as a function of time for 
various regions in the vicinity of Chernobyl. There is no 
evident increase in the CC following the accident.    
From the data in figure \ref{fig4} 95$\%$ confidence level upper limits 
on the increase in the CC in the month following the accident are 
9.3$\%$ (3.6$\%$) in the range 50-52.5N,30-32.5E(45-60 N, 20-35 E).\\ 
We estimate that the increase in ionization from this radioactivity  
relative to that produced by CR is a factor of 
$\sim15$ in the immediate vicinity of Chernobyl 
(50-52.5$^\circ$ N, 30-32.5$^\circ$ E) and a factor $\sim3$ in the 
fallout region 45-60$^\circ$ N, 20-35$^\circ$ E. The globally 
averaged modulation in the neutron 
monitor count rate is $8\%$. The modulation due to charged CR  
particles is roughly 1/3 of this. If the observed dip in CC of 
$\sim1\%$ \cite{MS1} is due to such modulation then the increase 
in CC due to the ionization from  Chernobyl should have  
been much greater than the upper limits quoted 
above and close to 100$\%$ CC anomaly.  
Hence the data from the  Chernobyl accident do  
not corroborate the claim of a causal connection between CR 
and CC.\\         
\section{Alternative explanation}
\begin{figure}
\begin{center}
\includegraphics [width=0.5\textwidth]{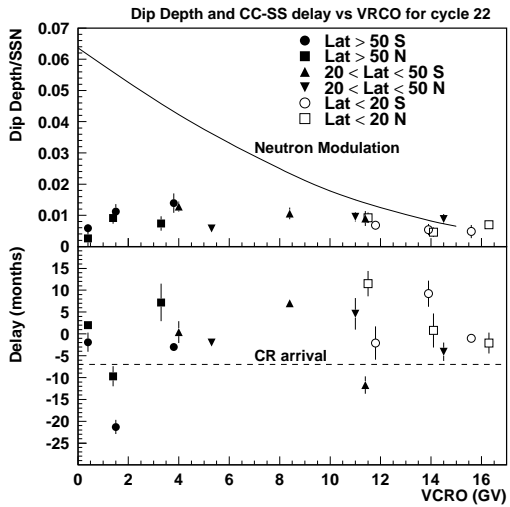}
\end{center}
\vspace*{-8mm}
\caption{The observed CC modulation (upper panel) as measured from the 
fit to solar cycle 22 only (see figure \ref{fig2}).   
The `modulations' are expressed by the dip amplitude per SS number 
at the time of solar maximum (1991). The smooth curve shows the 
modulation observed in neutron monitors around the World \cite{Braun} 
arbitrarily normalised.  The lower panel shows the fitted delay 
between the onset of the dip and that of the SS number in months. 
The dashed line shows the expected delay if a correlation existed 
between the changes in CR and CC. The delay between 
the CR and SS number increases is 7 months\cite{Kudela}.   
NB positive delay means CC preceeds the SS number increase.}\label{fig3}
\end{figure}

\begin{figure}
\begin{center}
\includegraphics [width=0.5\textwidth]{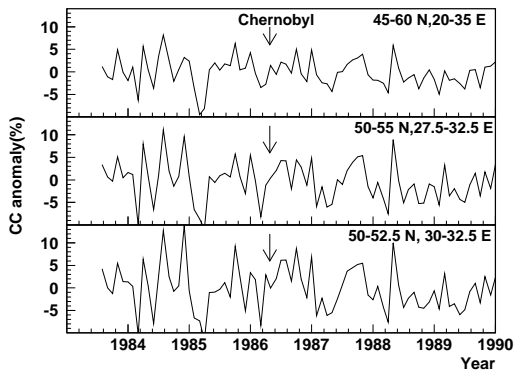}
\end{center}
\vspace*{-8mm}
\caption{CC against time in 3 regions in the vicinity 
of Chernobyl. The arrow shows the date of the nuclear accident. 
The upper panel is in the extended region where the fallout was greatest  
and the lower panels are smaller regions around Chernobyl.}\label{fig4}
\end{figure}

It seems likely that the origin of the solar
cycle modulation of the CC seen in Figures 1 and 2, and quantified
in Figure \ref{fig3}, is the 11-year cycle in solar irradiance, SI (e.g.
ref. \cite{Lean}) or possibly lightning.  The slow droop in CC with 
time, most evident
in Figure \ref{fig2} for the period after 1997, is possibly due to
anthropic factors although it has been suggested that this may be
due to instrumental drift \cite{Krist}. This
reduction is greater than adopted by us, in linear form, and
applied to the SS variation, suggesting an
accelerating `global warming'.\\
The contribution of SI to global warming is
estimated by us as having a peak to peak (SS cycle) magnitude of
about 10\% of the observed rise in mean global temperature over
the last 25 y.  The CR contribution appears to be much less than
10\%.\\
Finally, it is instructive to examine the energetics of the
various processes.  The ratio of energy input from the sun (SI) to
that from CR is $\sim 2.10^{8}$.  Of order 50\% of the CR energy
appears as ionization in the atmosphere, thus an `efficiency' as
low as $10^{-8}$ for SI in converting to ionization in the
atmosphere - or in other ways and resulting in cloud cover - is
all that is needed for SI to dominate.  Haigh \cite{Haigh} has, 
in fact,proposed such a mechanism involving solar-UV-ozone induced
dynamical feedback.
\section{Conclusions}
The dip in amplitude $1.3\%$ of the low altitude CC noted in 
reference \cite{MS1} in solar cycle 22 has been seen also in 
this analysis. This dip correlates well in amplitude and shape with 
the observed mean daily SS number.A number of attempts are  
described here to find evidence to corroborate the causal connection 
between the dip and changes in ionization levels due to CR as suggested 
in \cite{MS1}. 
The depth of the dip in solar cycle 22 is not a function of the 
increase in ionization as VRCO decreases. Nor is the onset of 
the dip well correlated with the arrival time of the increase in 
the CR rate. The dip 
in the following solar cycle, cycle 23, is only evident in the 
equatorial regions of the Earth. The atmospheric ionization 
produced by the nuclear accident at Chernobyl produced no 
observable increase in the CC. In summary, no corroboration 
of the claim of a causal connection between the changes in ionization 
from CR and CC was found in this investigation.  From the 
distribution of the change in solar cycle 22 with VRCO we find 
that less than 15$\%$ of the CC change comes from the CR 
modulation at 95$\%$ confidence level. In reference \cite{MS1} 
it is estimated that if all the CC change in cycle 22 were due to 
CR modulation 
then the radiative forcing produced by long term changes in the 
CR rate would be 1.4 Wm$^{-2}$ i.e. 
the major portion of that necessary to produce the global warming 
we observe.  
The upper limit on the effect of CR forcing of CC determined here 
implies that the radiative forcing due to CR is less than 
0.2 Wm$^{-2}$  at 95$\%$ confidence level. This can be used 
to set a limit on the global temperature rise from CR. 
Here we assume that $15\%$ of the observed dip in CC 
for cycle 22 is from the solar modulation of CR,  
taken to be $2.7\%$ for muons. We use the observation of a 
downward trend in the CR intensity \cite{Moraal} of $\sim 2\%$ in 
the last 35 years and the observation that the solar modulation 
of the average global temperature is $\sim0.1^\circ$C \cite{Lean}. 
From this the temperature rise due to CR must be less 
than $\sim 0.01^\circ$C.   

\end{document}